\documentclass[conference]{IEEEtran}
\IEEEoverridecommandlockouts
\usepackage{amsmath,amssymb,amsfonts}
\usepackage{algorithmic}
\usepackage{graphicx}

\usepackage{textcomp}
\usepackage[T1]{fontenc}
\usepackage{xcolor}
\usepackage[utf8]{inputenc}
\usepackage{csquotes}
\usepackage[english]{babel}
\usepackage[hidelinks,colorlinks=true,linkcolor=blue,citecolor=blue]{hyperref}

\DeclareMathAlphabet\mathrsfso      {U}{rsfso}{m}{n}
\def\BibTeX{{\rm B\kern-.05em{\sc i\kern-.025em b}\kern-.08em
    T\kern-.1667em\lower.7ex\hbox{E}\kern-.125emX}}

\usepackage[
backend=biber,
style=numeric,
sorting=none
]{biblatex}
\begin{document}
\title{Role of Modern Optical Techniques in Gravitational Wave Detection\\
}

\author{\IEEEauthorblockN{Parivesh Choudhary}
\IEEEauthorblockA{\textit{Department of Physics} \\
\textit{Indian Institute of Technology Kanpur}\\
\textit{Kanpur, India}\\
\href{mailto:parivesh@iitk.ac.in}{parivesh@iitk.ac.in}}\\
\textit{(Dated: December 2 2020)}
}

\maketitle
\thispagestyle{plain}
\begin{abstract}

Gravitational waves were first proposed by Henri Poincaré in 1905 and were subsequently predicted by Albert Einstein in his General Theory of Relativity. In 2015, first Gravitational Waves signals were detected by LIGO of two black holes merging about 1.3 billion light-years away. The discovery opened a new window of astronomy. Currently, four laser interferometers are operating around the world, two American detectors which are located in Hanford and Livingston, one Italian detector (VIRGO) located at Pisa and an underground Japanese detector KAGRA. This paper describes the modern optical techniques that are being used and future planned techniques for gravitational wave detection.    
\end{abstract}


\section{Introduction}
Just like Maxwell’s equation describe the relationship between electric charge and electromagnetic field, the Einstein field equations describe the interaction between mass and space-time curvature. Gravitational Waves (GW) are time de-pendent vacuum solutions to the field equations just like the electromagnetic waves are time dependent vacuum solutions to Maxwell’s equations. 

GW are a perturbation of the metric propagating in a Minkowski space-time. These are generally generated by rapidly accelerating large and compact masses, like merging black hole, spinning massive neutron stars, or supernovae explosions. When GW travel through space-time fabric, it compresses and stretches the fabric simultaneously (Fig.1). The effect of stretching and compressing can directly be observed by measuring the displacement between the two masses\cite{Kokkotas_2008,Calloni:2017whl}.

The induced displacement is directly proportional to the amplitude of the gravitational wave. The amplitude of the gravitation wave can be estimated as: 

\begin{equation}
h\sim\frac{G\epsilon E}{c^4 r}
\end{equation}

Where G is Gravitational constant, c is the speed of light, $\epsilon$ is the asymmetry of the source, E is a fraction of mass-produced and r is distance\cite{Kokkotas_2008}.

If for example, we take a supernovae explosion at Virgo cluster (15 Mpc) in which energy equivalent to 10${^{-4}}$M{\(_\odot\)} is produced, produces gravitational waves of 100 Hz and the signal duration is 1 msec, the amplitude of the gravitational wave on earth will be $\sim 10{^{-21}}$.

For a detector of length 4km, we are looking for the length difference of:

\begin{equation}
\Delta l = h\cdot l=10^{-22}\cdot 4km \sim 10^{-18}m
\end{equation}
which is 1000 times smaller than the diameter of a proton\cite{Abbott_2009}. The LIGO interferometers use special interferometry techniques, state-of-the-art optics, highly stable lasers and multiple layers of vibration isolation, to meet the design sensitivity.
\begin{figure}[htbp]
\centerline{\includegraphics[scale=0.5]{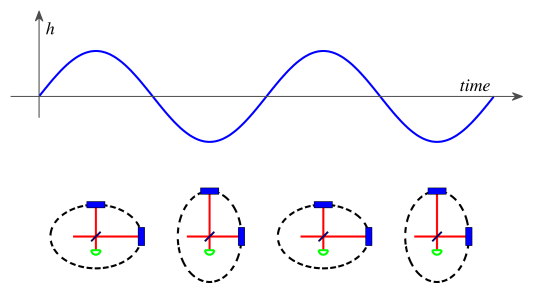}}
\caption{A strain amplitude h characterizes a GW travelling perpendicular to the plane of the diagram. The wave distorts the ring of particles into an ellipse, elongated in one direction in one half-cycle of the wave, and elongated in the orthogonal direction in the next half-cycle. This oscillation can be directly measured with a Michelson Interferometer as shown\cite{Abbott_2009}.}

\end{figure}

\section{Detector Description}
\subsection{Bar Detectors}
The first GW detector was built by an experimental physicist Joseph Weber at the University of Maryland in 1965. Weber used solid aluminium cylinders of size 2 meter long and 1 meter in diameter, which were suspended using steel wires. A passing GW would set these detectors into the resonant frequency of about 1660 Hz; they were just like tuning forks for detecting GW. Piezo-electric crystals were attached to these crystals to convert the vibrational energy of the bar to electrical signals\cite{Aguiar_2010}.

These bars were kept inside an isolated vacuum chamber to prevent EM interaction and acoustic noise interference. The two fundamental noise for these detectors were thermal noise which was caused by thermal vibration of aluminium atoms and electronic noise due to amplifiers which were used to amplify the electronic signals from piezo-electric crystals. 

By 1969, Weber thought that he might have detected GW and claimed many detections over the year. Many groups tried to replicate Weber's experiment, but, they didn't detect any signals Weber showed. The average amplitude of oscillation due to thermal oscillations were much larger than the oscillation amplitude of the GW.  

Ultimately, his claims were discarded due to many reasons, including that the experiment was not reproducible.

\subsection{Michelson Interferometer}
A Michelson Interferometer comprises of a laser source, beam splitter and two mirrors (Fig.2). Let the reflection coefficients of the mirrors be $r{_1}$ and $r{_2}$, placed at a distance of $l{_1}$ and $l{_2}$ from beam splitter. The incident light at the beam splitter is represented as\cite{Freise_2010}:
\begin{equation}
E_{i}(t) = E_{0}e^{i\omega t} 
\end{equation}

\begin{figure}[htbp]
\centerline{\includegraphics[scale=0.15]{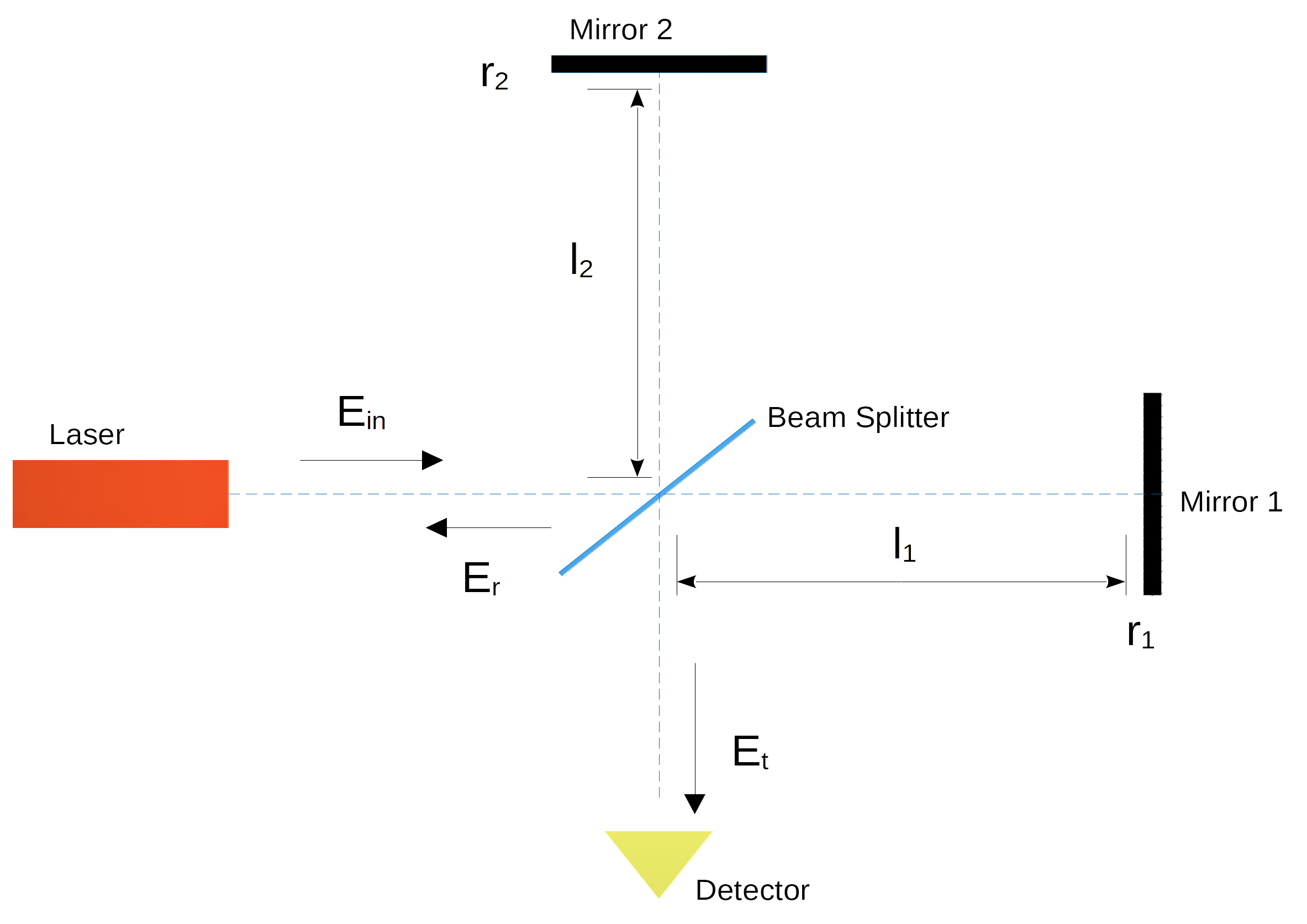}}
\caption{A simple Michelson Interferometer comprising of a beam splitter and two mirrors. The divided beams are recombined on the beam splitter.}

\end{figure}

The transmitted and reflected fields are related as:
\begin{equation}
E_{t} = E_{0}t_{b}r_{b}e^{i\omega t}(r_{1}e^{-i\theta_{1}}-r_{2}e^{-i\theta_{2}}) 
\end{equation}
\begin{equation}
E_{r} = E_{0}e^{i\omega t}(T_{b}r_{1}e^{-i\theta_{1}}+R_{b}r_{2}e^{-i\theta_{2}}) 
\end{equation}
where $\theta_{1}$, $\theta_{2}$, $t_{b}$, $r_{b}$, $T_{b}$, $R_{b}$ represents phase acquired by beams in the arms, transmission coefficient, reflection coefficient, transmittance, and the reflectance of the beam splitter respectively. 

The intensity of the fields is given by:
\begin{equation}
I_{t} = |E_{t}|^2 = I_{0}T_{b}R_{b}(R_{1}+R_{2}-2r_{1}r_{2}\cos{\Delta \theta}) 
\end{equation}
\begin{equation}
I_{r} = |E_{r}|^2 = I_{0}(T_{b}^{2}R_{1} + R_{b}^{2}R_{2}+2T_{b}R_{b}r_{1}r_{2}\cos{\Delta \theta})
\end{equation}
where $I_{0}$ is the intensity of the beam and $\Delta \theta$ is the phase difference of the two beams.

If we consider beam splitter and mirrors as test masses, the passing gravitational wave will cause the phase difference in two fields, thus showing interference at the output port. In other words, Michelson Interferometer (Fig.3) is sensitive to gravitational waves. This idea was originally developed by Rainer Weiss and Felix Pirani in 1967 and was further developed by Robert L. Forward.

\begin{figure}[htbp]
\centerline{\includegraphics[scale=0.7]{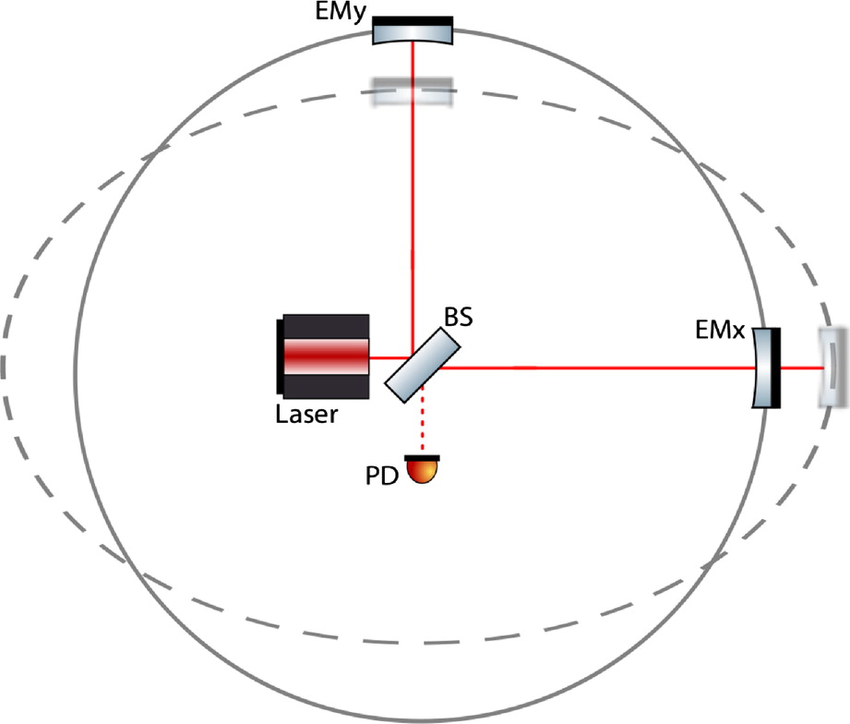}}
\caption{A Michelson Interferometer acting as a gravitational wave detector. When GW passes, one arm shortens while other lengthens, resulting in the interference pattern at the output port\cite{Hammond_2014}.}

\end{figure}

\subsection{Frequency Response of Michelson Interferometer}
The difference in phase of the two fields can be written as a sum\cite{article5,article6} of static phase($\Delta \theta_0$) and effect due to passing GW($\delta \theta$): 
\begin{equation}
\Delta \theta = \Delta \theta_0 + \delta \theta 
\end{equation}
\begin{equation}
\Delta \theta = \Delta \theta_0 + \underbrace{\int \Tilde{h}{(\Omega)}H_{MI}(\Omega,\omega)e^{i\Omega t}d\Omega}_\text{Fourier Transform of GW signal}
\end{equation}
Where $\Omega$ and $\omega$ are frequency of GW and beam respectively, $H_{MI}$($\Omega$,$\omega$) is the frequency response function given as:
\begin{equation}
H_{MI}(\Omega,\omega)=\frac{\omega}{\Omega}\Big(e^{-i\frac{\Omega l_{1}}{c}}\sin{\frac{\Omega l_{1}}{c}}+e^{-i\frac{\Omega l_{2}}{c}}\sin{\frac{\Omega l_{2}}{c}}\Big)
\end{equation}
neglecting higher order terms $H_{MI}$($\Omega$,$\omega$) can be approximated as:
\begin{equation}
H_{MI}(\Omega,\omega)\sim 2\frac{\omega}{\Omega}e^{-i\frac{\Omega \Tilde{l}}{c}}\sin{\frac{\Omega \Tilde{l}}{c}}
\end{equation}
\begin{equation}
\Tilde{l}=\frac{l_{1}+l_{2}}{2}    
\end{equation}
Taking $|H_{MI}(\Omega,\omega)|$ maximum, therefore:
\begin{equation}
l=\frac{(2n+1)c\pi}{2\Omega}    
\end{equation}
\begin{equation}
l=\frac{c}{4f}
\end{equation}
Thus, for example, we take a sinusoidal GW with a frequency of 500Hz, the length of arm required to detect is 150km!!
\subsection{Fabry-Perot Interferometer}
As described in the previous section, the optimum length of the Michelson Interferometer arm to detect GW from astronomical sources is the order of 100km. However, this is unrealistic to build such a large detector on the surface of Earth due to current technological limitations. Therefore, to overcome this limitation, the optical paths length of the interferometer are folded by using the Fabry-Perot cavity.    

A Fabry-Perot cavity consists of two partial reflective glass having a spacing between them. When light enters the cavity, it is partially transmitted from the first mirror, reflected multiple times inside the mirror, and then transmitted back again by the front mirror. The multiple reflections inside the cavity make the effective optical path length of light much larger than the length of the cavity\cite{article6}.

Let us consider two mirrors placed parallel to each other at a distance of L. A plane wave of frequency $\omega$ and wavenumber k travelling in the z-direction is expressed as:
\begin{equation}
E_i{(t,z)}=E_0e^{i(\omega t-kz)}    
\end{equation}

\begin{figure}[htbp]
\centerline{\includegraphics[scale=0.17]{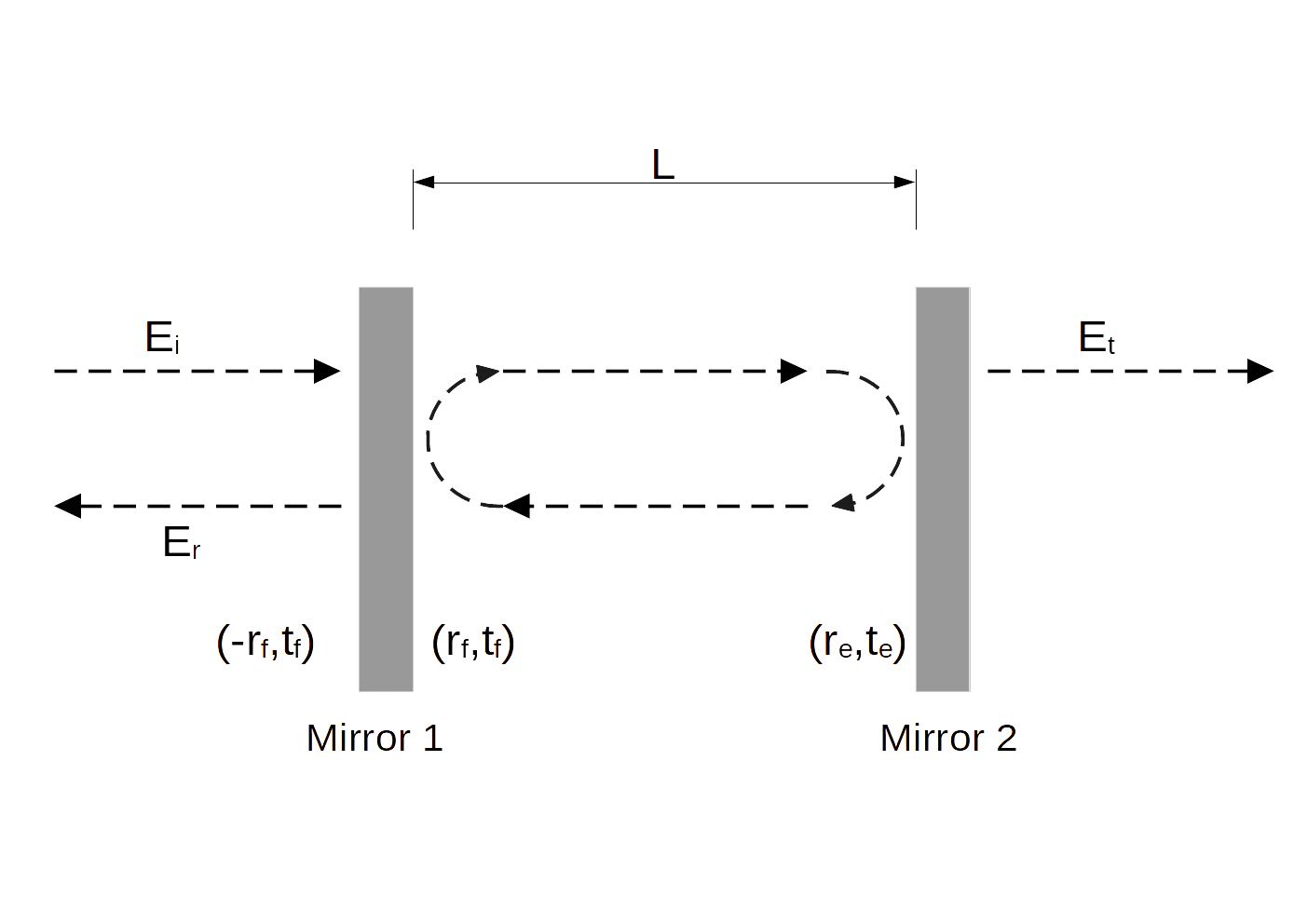}}
\caption{A simple Fabry Perot cavity. Two mirrors placed at a distance of L. Let mirror 1 be the front mirror and mirror 2 be end mirror.}

\end{figure}
The reflection and transmission coefficients of mirrors are represented as ($r_f$,$t_f$) and ($r_e$,$t_e$). The reflected and transmitted light outside the cavity is expressed as:
\begin{equation*}
\begin{aligned}
E_r=-E_0r_fe^{i(\omega t+kz)}+ &\\  E_0r_eT_fe^{-i\Phi_0}\sum_{n=0}^{\infty}&(r_fr_ee^{-i\Phi_0})^n e^{i(\omega t+kz)} 
\end{aligned}
\end{equation*}
\begin{equation*}
=E_0\bigg[-r_f+r_eT_fe^{-i\Phi_0}\sum_{n=0}^{\infty} (r_fr_e e^{-i\Phi_0})^n\bigg]e^{i(\omega t+kz)}
\end{equation*}
\begin{equation}
E_r=E_0\bigg[-r_f+r_eT_f\frac{e^{-i\Phi_0}}{1-R_{fe}e^{-i\Phi_0}}\bigg]e^{i(\omega t+kz)} 
\end{equation}
\begin{equation*}
E_t=E_0t_ft_e\sum_{0}^{\infty}(r_fr_e e^{-i\Phi_0})^n e^{(i\omega t-kz)}    
\end{equation*}
\begin{equation}
=E_0\frac{T_{fe}}{1-R_{fe} e^{-i\Phi_0}}e^{(i\omega t-kz)}    
\end{equation}
where $R_i=r{_i}^2$, $T_i=t{_i}^2$ ( for i=f,e), $R_{fe}=r_f r_e$, $T_{fe}=t_f t_e$ and $\Phi_0 = \frac{2\omega L}{c}$ is the round trip phase of light inside the cavity. The reflection and transmission coefficients can be calculated as:
\begin{equation}
r_c(\Phi_0)=-r_f + r_eT_f\frac{e^{-i\Phi_0}}{1-R_{fe}e^{-i\Phi_0}}    
\end{equation}
\begin{equation}
t_c(\Phi_0)=\frac{T_{fe}}{1-R_{fe}e^{-i\Phi_0}}e^{-i(\frac{\Phi_0}{2})} \end{equation}
The Finesse ($\mathrsfso{F}$) of a Fabry Perot interferometer is related to reflectivity of mirror as:
\begin{equation}
\mathrsfso{F} = \frac{\pi\sqrt{R_{fe}}}{1-R_{fe}}
\end{equation}
The Finesse ($\mathrsfso{F}$) is also related to the number of round trips the light makes inside the cavity as:
\begin{equation}
N=\frac{2\mathrsfso{F}}{\pi}    
\end{equation}
Taking $\Phi_0 \ll 1$ and $\mathrsfso{F} \gg 1$, reflection and transmission coefficient can be approximated as:
\begin{equation*}
r_c(\Phi_0)\sim -r_f+T_f\frac{\mathrsfso{F}}{\pi + i\mathrsfso{F}\Phi_0}    
\end{equation*}
\begin{equation*}
= -r_f+T_f\frac{\mathrsfso{F}}{\pi}\frac{1}{1 + i\frac{\mathrsfso{F}}{\pi}\Phi_0}    
\end{equation*}
\begin{equation}
= -r_f+T_f\frac{\mathrsfso{F}}{\pi}\frac{1}{1 + i\frac{\omega}{\omega_c}}    
\end{equation}
\begin{equation*}
t_c(\Phi_0)\sim T_{fe}\frac{\mathrsfso{F}}{\pi}\frac{1}{1 + i\frac{\mathrsfso{F}}{\pi}}    
\end{equation*}
\begin{equation}
=T_{fe}\frac{\mathrsfso{F}}{\pi}\frac{1}{1 + i\frac{\omega}{\omega_c}}    
\end{equation}
where ${\omega_c}^{-1}$ is the storage time, given as:
\begin{equation}
 \omega_c^{-1}=\frac{2\mathrsfso{F}L}{c\pi}  
\end{equation}

\subsection{Frequency Response of Fabry-Perot Interferometer}
When GW passes, it changes the proper length of the cavity, thus affects the round trip phase of the light. Let $\Delta t_n$ be the delay of the wave which arrives at the first mirror after n round trips, therefore reflected field at the first mirror is\cite{article6}:
\begin{equation}
E_r=E_0e^{i \omega t}\big[-r_f+r_eT_f\sum_{1}^{\infty}(r_fr_e)^{n-1}e^{(-i\omega\Delta t_n)}\big]    
\end{equation}
\begin{equation}
E_r=E_0e^{i\omega t}r_c{(\omega)}\bigg[1-i\int \Tilde{h}{(\Omega)}H_{FP}(\Omega,\omega)e^{i\Omega t}d\Omega\bigg] 
\end{equation}
where $H_{FP}(\Omega , \omega)$ is the response the function of the Fabry-Perot cavity. When the frequency of light is tuned to the resonance of the cavity i.e. $\frac{\omega l}{c}=n\pi$, response function is given as:
\begin{equation}
H_{FP}\sim \frac{\omega}{2\omega_c}\frac{T_f\mathrsfso{F}}{\pi r_c{(0)}}\frac{1}{1+i\frac{\Omega}{\omega_c}}
\end{equation}
The phase difference can be approximated as:
\begin{equation}
\delta \theta = \frac{8l\mathrsfso{F}}{\lambda}\frac{he^{i(\Omega t)}}{1+i\Omega \frac{2l\mathrsfso{F}}{\pi c}}
\end{equation}
\begin{figure}[htbp]
\centerline{\includegraphics[scale=0.5]{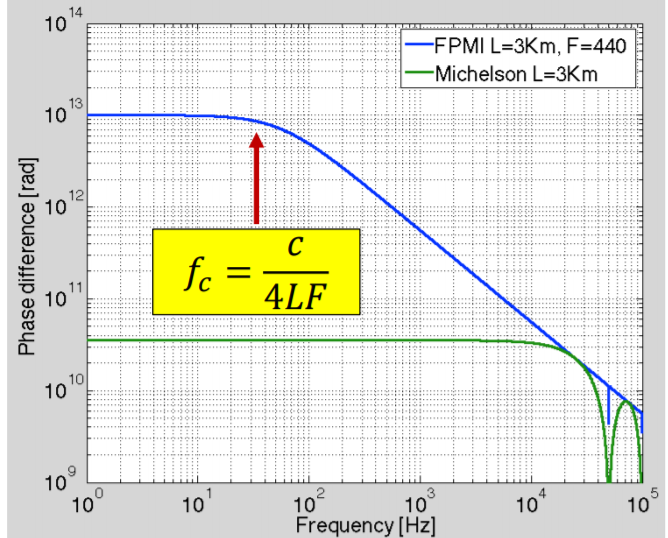}}
\caption{Comparison between the frequency response of Michelson Interferometer and Fabry-Perot cavity. For advanced VIRGO $\mathrsfso{F}=440$, $N=280$, $l=3km$; the effective arm length increases by the factor of N\cite{Bertolini:2017}.}

\end{figure}

\section{Laser and sources of noise}
One key technology necessary to reach the design sensitivity of these detectors is ultra-stable high power laser. The fundamental noise in the LIGO laser system is readout noise, and it is the only fundamental noise due to laser. The readout noise is the combination of shot noise and pressure noise. 

\subsection{Shot noise}
The shot noise arises due to random occurrences of photons on the detector. To measure the power of light is the same thing to count the number of photons in a time unit. The random occurrences can be modelled as  Poisson distribution. The shot noise on the cavity is given as:
\begin{equation}
h_{sh}(f)=\frac{\Delta l}{l}=\frac{1}{l}\sqrt{\frac{\hbar c \lambda}{2\pi T(f)P_{in}}}
\end{equation}
where $P_{in}$ is the input power, l is the length of the interferometer, $\lambda$ is the wavelength of the input light and T(f)=d($P_{out}$/$P_{in}$)/d$\phi$ is the unit-less transfer function of the interferometer. The shot noise is inversely proportional to input power and directly proportional to wavelength. Using high power lasers, shot noise can be minimized. The choice of wavelength cannot be arbitrary since high power laser devices are not available at all wavelengths. Currently, LIGO uses Nd:YAG lasers, with the wavelength of 1064nm. These are industrial grade lasers capable of producing a continuous wave of 100's of Watts. 

\subsection{Pressure noise}
The pressure noise is due to pressure applied by photons on the mirrors. It is given as:
\begin{equation}
h_{rp}(f)=\frac{\Delta l}{l}=\frac{2}{mf^2}\sqrt{\frac{\hbar T(f)P_{in}}{8\pi^3 c\lambda}}
\end{equation}
The pressure noise increases when the power of the laser increases, and it is
high at low frequency. The total readout noise is:
\begin{equation}
h_{ro}=\sqrt{h_{sh}^2 + h_{rp}^2}    
\end{equation}
Thus it is necessary to choose the right power of laser which gives minimum $h_{ro}$ at the particular frequency\cite{article4}.

\subsection{Pre-stabilized laser}
The pre-stabilized laser (PSL) of the LIGO detectors is developed by the Laser Zentrum, Albert-Einstein-Institute (AEI) and neoLASE in Hannover, Germany. The laser system consists of three stages. It is enclosed in a class 1000 clean-room.  It is very important that the laser is in a clean environment as particles in the beam path and on optical components lead to scattering and thus a reduction in the beam quality. 

\subsubsection{Stage 1}
The first stage consists of the non-planar ring oscillator (NPRO Fig.6). This solid-state laser uses an Nd:YAG (neodymium-doped yttrium aluminium garnet) crystal (of size about 1cm) as the laser medium. The front surface of the crystal is a partially reflective dielectric coating, which is highly transmissive to the input light and also serves as the mirror of the resonator. On all other surfaces, total internal reflection occurs. The NPRO is pumped by a laser diode having wavelength of 808nm and delivers a single-frequency output power of 2W\cite{phdthesis,article1,Pold:2014qzc}.  

\begin{figure}[htbp]
\centerline{\includegraphics[scale=0.50]{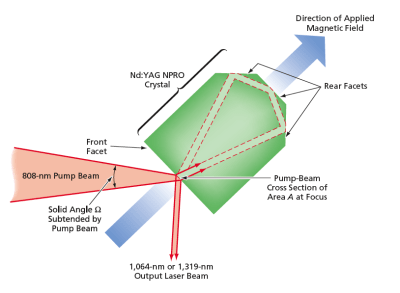}}
\caption{Setup of a non-planar ring oscillator laser. The front face of the crystal has a dielectric coating, serving as the output coupler and also a partially polarizing element, facilitating unidirectional oscillation. The red beam is the pump beam, generated with a laser diode of 808nm\cite{Liu:2007}.}

\end{figure}

\begin{figure}[htbp]
\centerline{\includegraphics[scale=0.35]{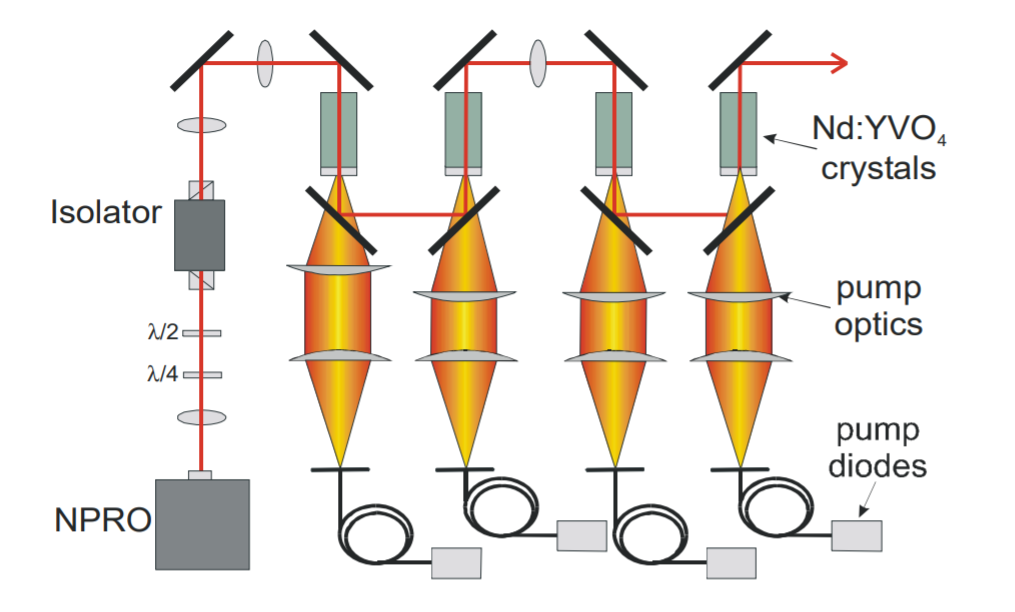}}
\caption{Setup of the four stage amplifier design with an NPRO seed source\cite{Frede:07}.}

\end{figure}

\subsubsection{Stage 2}
The second stage consists of a single-pass amplifier having an output power of 35W. The laser from NPRO is sent to four Nd:YVO$_{4}$ crystals (Fig. 7). Each laser material is pumped by a 400 $\mu$m diameter fibre-coupled laser diodes delivering a maximum output power of 45 W. To separate pump light from input light, a dichroic 45$^\circ$ mirror with anti-reflection coating for the pump wavelength and high-reflection coating for the laser wavelength is used\cite{article}. 95$\%$ of the output power is within TEM$_{00}$ mode.   

\subsubsection{Stage 3}
The third stage consists of a rod-shaped device called High Power Oscillator (HPO). The beam is made to pass through four bundles of fibre containing seven fibres each, which are arranged in a hexagonal pattern. Each fibre carries 45W power, therefore each bundle gives theoretical power of 315W. By the time beam leaves the oscillator, its power reaches around 200W. The frequency of the output beam is matched to the output of NPRO using series of loop controls, which results in ultra-stable frequency laser.

\begin{figure}[htbp]
\centerline{\includegraphics[scale=0.38]{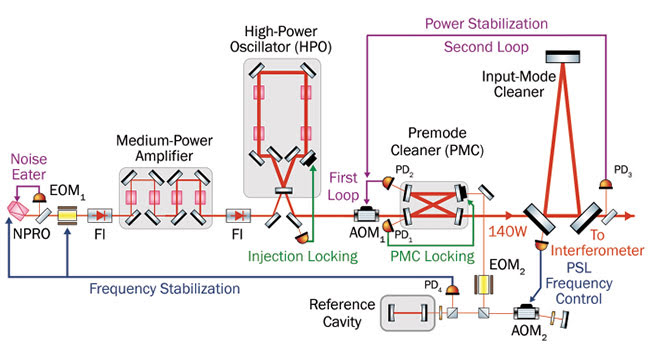}}
\caption{Pre-stabilized laser system of Advanced LIGO. The three-staged laser: NPRO, medium power amplifier and high power oscillator\cite{article3}.}

\end{figure}

\section{Power Recycling}
\subsection{Principle of power recycling}

One of the fundamental noise which limits the sensitivity of the interferometer  is shot noise. As previously seen, shot noise is inversely proportional to the square root of the incident power.  Since the gravitational wave signal is proportional to the incident power, the shot-noise level of an interferometer is improved with high laser power.

Besides increasing the laser power, shot noise can also be improved by a technique called Power Recycling. In order to decrease the shot noise, the interferometer is operated in the dark fringe condition, i.e. the reflected beams from the arms are made to interfere destructively at the output port. In this case, the injected laser beam is almost reflected back to the laser source. In this technique, the reflected laser beam is made to reflect back to the interferometer by keeping a partially transmitting mirror (called power recycling mirror) between laser source and interferometer, which results in the increase of power inside the interferometer\cite{Izumi:203yfa}.  

\subsection{Recycling cavity}

The additional cavity which is formed due to this recycling mirror is called Recycling cavity\cite{phdthesis4}. Let's consider a recycling cavity (Fig. 9) formed by keeping mirror (PR; for simplicity, assume only one recycling mirror is kept). The fields are given as:
\begin{equation}
E_{t}=t_{R}E_{i}+r_{R}E_{circ}
\end{equation}
\begin{equation}
E_{circ}=r_{com}E_{t}
\end{equation}
\begin{equation}
E_{r}=-r_{R}E_{i}+t_{R}E_{circ}
\end{equation}
where $E_{i}$, $E_{t}$, $E_{circ}$, $t_{R}$, $r_{R}$, $r_{com}$  is the incident light, transmitted light, circulated light inside the cavity, transmission coefficient, reflection coefficient of recycling mirror and complex reflectivity of the entire mirrors of the Fabry-Perot cavity. From the above equations, we obtain the following expressions:
\begin{equation}
g=\frac{E_{t}}{E_{i}}=\frac{t_{R}}{1-r_{R}r_{com}}
\end{equation}
\begin{equation}
r_{rec}=\frac{E_{r}}{E_{i}}=-r_{R}+ \frac{t_{R}^2 r_{com}}{1-r_{R}r_{com}}
\end{equation}
where g and r$_{rec}$ is the recycling cavity  amplitude gain and reflectivity of the recycling cavity, respectively. The incident beam resonates with the recycling cavity when r$_{com}$ is a real and positive number. 

\begin{figure}[htbp]
\centerline{\includegraphics[scale=0.15]{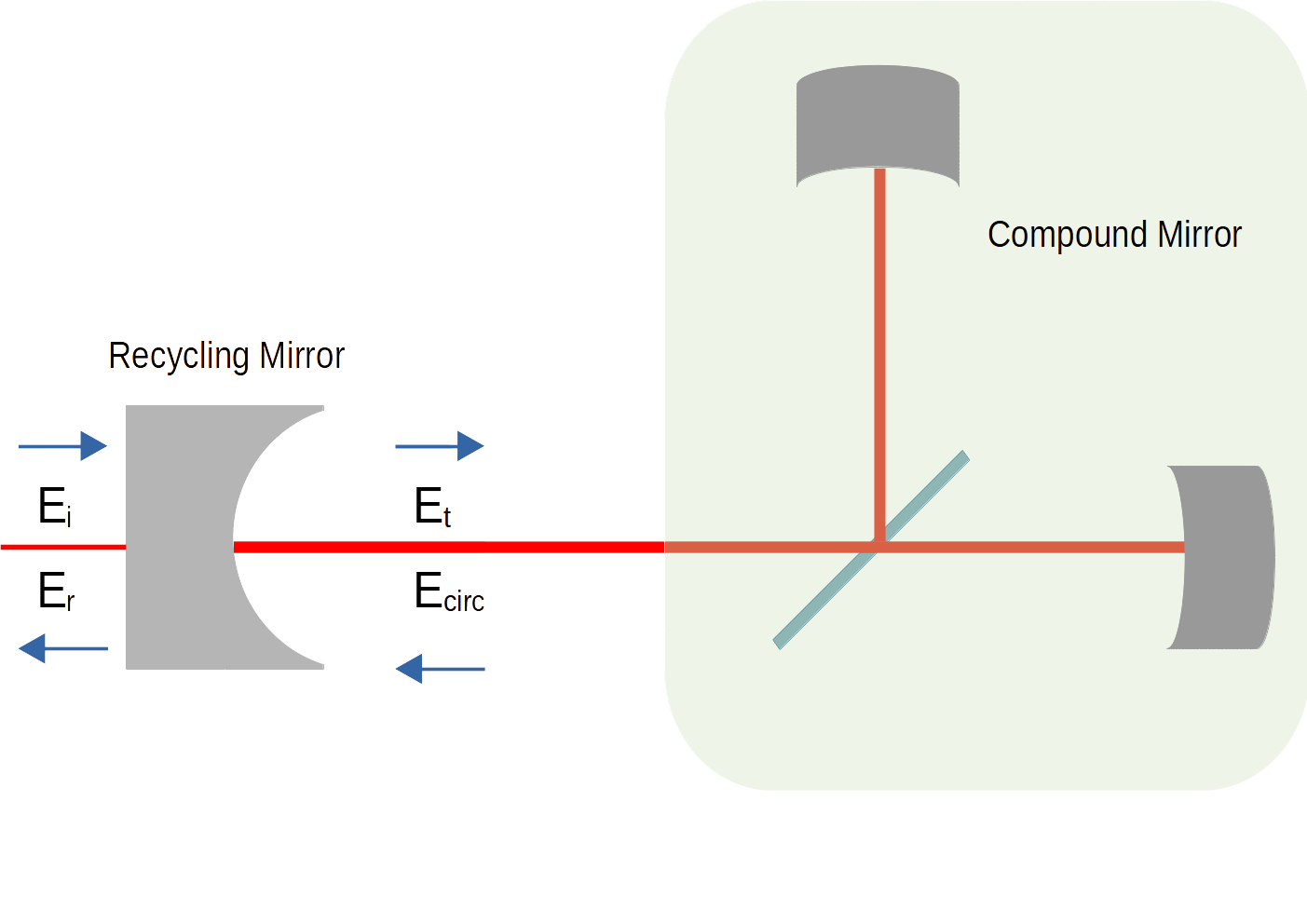}}
\caption{Recycling cavity formed by a recycling mirror and a compound mirror.}

\end{figure}

\subsection{Power recycling gain}

Power recycling gain or a power recycling factor (G) is the ratio of the laser power on the beamsplitter with and without the recycling mirror, which is given as:
\begin{equation}
G=g^2=\bigg( \frac{t_{R}}{1-r_{R}r_{com}}\bigg)^2
\end{equation}
When G>1, the effective power of the laser source increases to GP and the shot noise is reduced to the factor of $\sqrt{G}$. The typical power recycling gain expected for the LIGO detector is between 30 and 50. The power recycling gain is maximized when the reflectivity of the recycling mirror is equal to the reflectivity of the compound mirror i.e.\cite{Izumi:203yfa}:
\begin{equation}
|r_{r}|^2=|r_{com}|^2
\end{equation}

\begin{figure}[htbp]
\centerline{\includegraphics[scale=0.48]{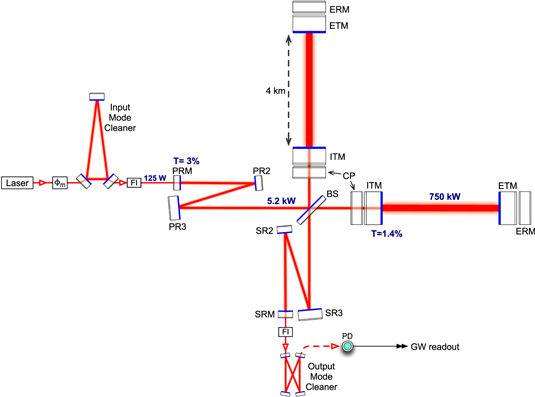}}
\caption{Advanced LIGO optical configuration.  Where ITM: input test mass; ETM: end test mass; ERM: end reaction mass; PR2/PR3: power recycling mirror 2/3; PRM: power recycling mirror;  CP: compensation plate; BS: 50/50 beam splitter; SRM: signal recycling mirror; SR2/SR3: signal recycling mirror 2/3; FI: Faraday isolator;  PD: photodetector, $\phi_m$: phase modulator. All of the components shown are mounted in the LIGO ultra-high vacuum system on seismically isolated platforms\cite{2015}.}

\end{figure}

\section{Future}

\subsection{Space based interferometers}

Currently, LIGO has one of the largest vacuum systems on Earth; it takes 40 days of constant pumping to make an ideal operating pressure. Creating such a large vacuum is one of the current technological limitations. To overcome this, space based interferometer are planned.

One of the future planned space GW detector is the Laser Interferometer Space Antenna (LISA) jointly planned by ESA and NASA. In contrast to ground-based GW detectors that have a typical sensitivity in the range from 1Hz to 1kHz, the sensitivity for LISA stretches between 0.1mHz and 0.1Hz. 

\begin{figure}[htbp]
\centerline{\includegraphics[scale=0.28]{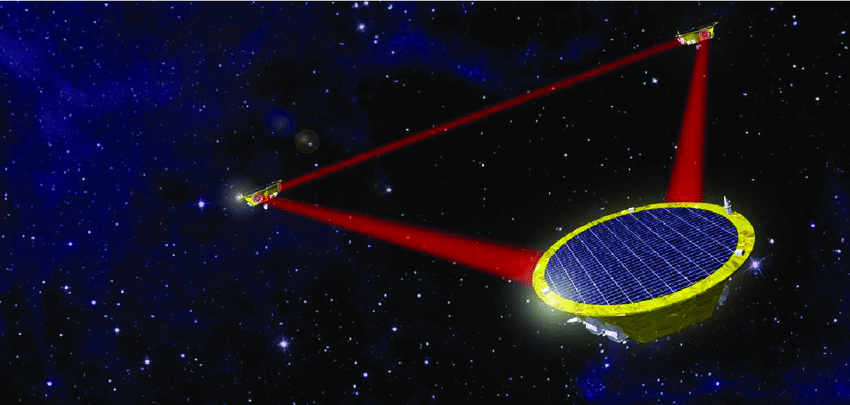}}
\caption{Artist's view of the LISA space mission. Three satellites form a Michelson interferometer with an arm-length of $\sim$ 2.5 million kilometers\cite{phdthesis3}.}

\end{figure}

LISA comprises of 3 spacecraft in a heliocentric orbit, forming an equilateral triangle with a side of 2.5 million kilometres (Fig. 11). Each spacecraft has two test masses, kept in a free-fall condition; they form the reference points for interferometric measurement of the inter spacecraft distance. To measure the distance between two spacecraft, a 2W laser beam of 1064nm is sent through a 40cm Cassegrain telescope. The telescope receives and transmits the laser beam at the same time. In an ordinary interferometer, the received light from another spacecraft is sent back where the light would combine with the local oscillator to complete the measurement. 

However, due to the large distance between the spacecraft, this method is not feasible. Due to diffraction, the transmitted laser beam widens to many kilometres at the receiving spacecraft; the received power is the order of few picowatts\cite{Jennrich_2009,article7}.

In the LISA satellites, a transponder scheme with offset phase-locking is used. The received light is combined with a local oscillator derived from the transmitting laser, and the phase difference is measured. The laser beam is sent to the third spacecraft; the phase of the laser beam is the exact copy of the received laser light. The GW signal is calculated by measuring the phase and receiving time of each spacecraft\cite{Danzmann_1996}.

\subsection{Atom interferometery}
Apart from shot noise and pressure noise, the sensitivity of ground detectors is greatly limited due to seismic noise. These noises are generally caused due to ground vibrations, wind, human activities such as logging or trains, ocean waves, etc. The sensitivity of current ground Laser Interferometers at frequencies below $\sim$ 40Hz is poor because of this noise. 

The GW spectrum between 10 Hz and 10$^{-3}$ Hz has very exciting sources. GW emerging from sources such as white dwarf binaries, and intermediate and massive black holes occur in this band. This band is also interesting for the search for stochastic GW searches\cite{Dimopoulos_2008}.

In an atom interferometer wave character of atoms is unlike photons. Just like laser interferometers, an atom is forced to follow a superposition of two spatially and temporally separated free-fall paths. The atom is coherently split using pulses of light which transfers momentum to a part of the atom. When this atom is combined later, it shows interference due to phase accumulated in these paths. Generally, this follows the Mach-Zehnder interferometer configuration. An atom interferometer consists of 3 stages: atom cloud preparation, interferometer pulse sequence, and detection.

\subsubsection{Stage 1}
In the first stage, a sub-microkelvin cloud of atoms is formed using laser cooling. The basic principle of the laser cooling is that if an atom is travelling towards a laser and it absorbs the photon, it will be slowed down. The frequency of the laser is chosen to be somewhat higher than the atomic resonance. This stage is important because many systematic errors are sensitive to the initial condition of the atoms, so cooling mitigates these unwanted errors. At the end of the cooling stage, the final atom cloud has a very low density, which is enough to neglect the atom-atom interaction. This cloud is then launched with a velocity(v$_{l}$) by transferring momentum using laser light.

\subsubsection{Stage 2}
In the second stage, the atom cloud is made to follow a trajectory. A sequence of light pulses is used to steer these atoms (Fig. 11). These pulses serve as beam splitters and mirrors that coherently divide each atom’s wave-packet and then later recombine it to produce the interference. In an atom beamsplitter, laser light having wavevector k$_1$ is initially absorbed by the atom. Subsequently, another laser light having wavevector k$_2$ is made to strike atom, which stimulates the emission of a photon of wavevector k$_2$. This results in a momentum transfer of:
\begin{equation}
k_{eff}=k_{2}-k_{1}\sim 2k_{2}
\end{equation}
The laser lights causes Rabi oscillations between states $|p>$ and $|p+\hbar k_{eff}>$ states. When laser pulse time is equal to a quarter of a Rabi period ($\frac{\pi}{2}$), it results in beamsplitter, and when laser pulse time is equal to the half of a Rabi period ($\pi$), it results in a mirror. After the action of first beamsplitter pulse, the atom is in superposition state which differ in velocity by $\frac{k_{eff}}{m}$. This spatial difference of atom is proportional to the sensitivity of the interferometer towards the GW strain. A final beamsplitter pulse is applied to combine these two components, which results in interference. 

\begin{figure}[htbp]
\centerline{\includegraphics[scale=0.30]{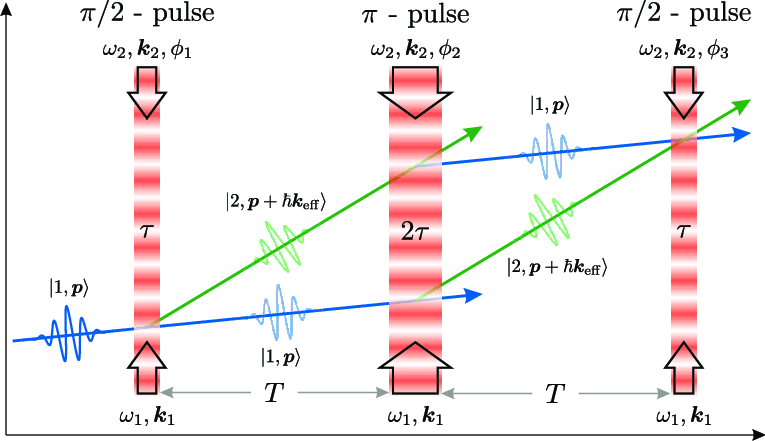}}
\caption{Three-pulse atom interferometer. Here,${k_{eff}}$ = $k_{2}-k_{1}$ is the effective wavevector for the two-photon transition\cite{article8}.}

\end{figure}

\subsubsection{Stage 3}
The third stage is atom detection. At the end of the second beamsplitter pulse, each atom is in a superposition of the two output velocity states which differ by $\frac{k_{eff}}{m}$. After an appropriate drift time, the two states are resolved separately and measured using absorption imaging. The total phase difference can be approximated as\cite{Dimopoulos_2008}:
\begin{equation}
\Delta \phi \propto hk_{2}\big(x_{i}-\frac{D}{2}\big)
\end{equation}
where h, k$_2$, x$_{i}$ and D are GW strain, the wavevector of the second laser, distance between atom and second laser, and distance between laser respectively.

The current challenge in this application is to cool the atom to the required narrow velocity distribution and support high repetition rate. On earth, the repetition rate of 10Hz is required for the GW band of 1-10Hz. Low-density cloud is required to mitigate the systematic error associated with cold atom collisions.

\section{References}
\printbibliography[heading=none]

\end{document}